\documentclass[aps,pra,twocolumn,showpacs,groupedaddress,letterpaper]{revtex4-1}

\usepackage{graphicx}  
\usepackage{float}
\usepackage{dcolumn}   
\usepackage{bm}        

\usepackage{float}
\usepackage{graphicx} 
\usepackage{array}    
\usepackage{amsmath}   
\usepackage{amssymb}   
\usepackage{wrapfig}  

\usepackage{color}  
\usepackage{setspace}
\usepackage[flushleft]{threeparttable}

\begin{document}
\widetext
\title{Magneto-Optical Trapping of Holmium Atoms}
\author{J. Miao, J. Hostetter, G. Stratis, and M. Saffman}
\affiliation{
 Department of Physics, University of Wisconsin,
1150 University Avenue, Madison, Wisconsin 53706
 }

\date{\today}
														
\begin{abstract}
We demonstrate sub-Doppler laser cooling and magneto-optical trapping of the rare earth element Holmium. Atoms are loaded from an atomic beam source and captured in six-beam $\sigma_+ - \sigma_-$ molasses using a strong $J=15/2 \leftrightarrow J=17/2$ cycling transition at $\lambda=410.5~\rm  nm$. Due to the small difference in hyperfine splittings and Land\'e $g$-factors in the lower and upper levels of the cooling transition the 
MOT is self-repumped without additional repump light, and deep sub-Doppler cooling is achieved with the magnetic trap turned on. We measure the leakage out of the cycling transition to metastable states and find a branching ratio $< 10^{-5}$ which is adequate for state resolved measurements on hyperfine encoded qubits.  
\end{abstract}

\maketitle

The magneto-optical trap (MOT) is a standard and very widely used tool in cold atom physics. To date MOT operation has been demonstrated for about 30  different neutral elements. Operation of a MOT on an open transition which allows for pumping into metastable states which are not cooled requires ``repump" light to return atomic population to the levels participating in laser cooling. The rare earth lanthanides did not originally appear amenable to laser cooling due to the presence of an 
open $f$ shell (except for Yb which was laser cooled in 1999\cite{Honda1999}), and  a  correspondingly complex atomic structure. It was recognized by McClelland that  the  lanthanides  are amenable to laser cooling due to the presence of cycling transitions between odd(even) parity ground states with angular momentum $J$ and even(odd) parity excited states  with angular momentum $J'=J+1$. MOT operation was first demonstrated with an open $f$ shell lanthanide in 2006 using Er\cite{McClelland2006a}. That demonstration was followed in 2010 by demonstration of  
MOTs for Dy\cite{LuYoun2010} and Tm\cite{Sukachev2010}. In this work we report on the operation of a Ho MOT loaded from an atomic beam with and without laser slowing\cite{Miao2013a}. The Doyle group has also recently demonstrated lanthanide MOTs including Ho  using buffer gas precooling of an atomic beam \cite{Hemmerling2013}.

Interest in laser cooling and trapping of lanthanide atoms is motivated by several topics in current research. Quantum degenerate bosonic and fermionic gases of 
Dy\cite{LuBurdick2011,LuBurdick2012}, and Er\cite{Aikawa2012,Aikawa2013} open up new research directions due to the large magnetic moments Er $(7~\mu_{\rm B}),$ Dy $(10~\mu_{\rm B})$   which provide for much stronger dipolar magnetic  interactions between neutral atoms  than are present in the more widely studied alkali gases.  
Ho has one stable isotope $^{165}$Ho, which is bosonic with nuclear spin of $I_{\rm nuc} = 7/2$ and a ground electronic 
configuration [Xe]$4f^{11}6s^2$. The ground state term is $^{4}I^{\rm o}, J = \frac{15}{2}$ giving eight hyperfine levels $F_g=4...11$, and the 
  magnetic moment is $9~\mu_{\rm B}$. Ho is distinguished by having 128 hyperfine-Zeeman states, the largest number  of any stable atomic isotope. This large number of states is of interest for collective encoding of multi-qubit quantum 
registers\cite{Brion2007d,Saffman2008}. Implementation of collective encoding will rely on Rydberg blockade interactions\cite{Saffman2010} in a dense atomic sample. The demonstration  of sub-Doppler laser cooling, MOT operation, and optical cycling transitions with fractional leakage rates at the $10^{-5}$ level  
reported here are  first steps towards Rydberg spectroscopy and qubit encoding in Ho atoms. 

\begin{figure}[!t]
 \centering
   \includegraphics[width=0.5\textwidth]{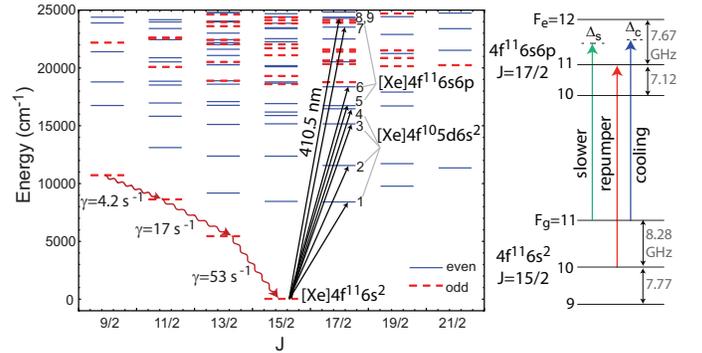}
  \vspace{-.8cm} \caption{(color online) Energy levels of Ho. The upper levels of the cycling  and cooling transitions are labeled 1-7. The diagram on the right shows the cooling, repumper, and slowing light used on the 410.5 nm transition to level $|e\rangle$.  }
   \label{fig.Holevels}
\end{figure}

\begin{table}[!t]
\begin{threeparttable}
\caption{
\label{tab.Hocooling} Ho cycling and cooling transitions. The columns list the vacuum wavelength, natural linewidth, and Doppler cooling limit $(T_{\rm D}=\hbar\gamma/2k_B)$.} 
\begin{ruledtabular}
\begin{tabular}{cccc}
 Transition&$\lambda ({\rm nm})$& $\gamma/2\pi({\rm MHz})$ & Doppler limit($\mu\rm K$)\\ \hline
    1 & 1193. & unknown & --\\
    2 & 867.3 & unknown & --\\ 
    3 & 660.9 & unknown & --\\ 
 4 & 608.3 & 0.038$^{\rm a}$& 0.91\\ 
 5 & 598.5 & 0.146$^{\rm b,c}$ & 3.5\\ 
 6 & 545.3 & unknown & --\\ 
7 & 425.6 & 1.59$^{\rm b,c}$ &38.\\
8$^{\rm d}$ & 412.1 & 2.3$^{\rm b,c}$ &55.\\
	9 & 410.5 & 32.5$^{\rm b,c}$ & 780.\\ 
\end{tabular}
\end{ruledtabular}
\noindent
\begin{tablenotes}
\item Wavelengths are calculated  from  \cite{Kramida2013}.
\item $^{\rm a}$ Linewidth derived from oscillator strength value reported in  \cite{Gorshkov1979}, $^{\rm b}$ Ref. \cite{DenHartog1999},
 $^{\rm c}$  Ref. \cite{Nave2003}, $^{\rm d}$ Two electron jump transition.
\end{tablenotes}
\end{threeparttable}
\end{table}

The level diagram of Ho together with   cycling transitions suitable for cooling is shown in Fig. \ref{fig.Holevels}. The ground state has $J=15/2$ with odd parity. 
There are eight dipole allowed transitions at wavelengths longer than 410 nm  to levels with even parity and $J=17/2$. Since there are no odd parity levels with $15/2\le J\le 19/2$ 
in between the ground state and the upper levels of the six transitions labeled 1-6  these transitions are cycling. Relevant transition parameters are given in Table \ref{tab.Hocooling}. The 410.5 nm  transition from the ground state to [Xe]$4f^{11}6s6p(^1P), J=17/2$ at 24361 cm$^{-1}$ (labeled 9 in Fig. \ref{fig.Holevels}) is not cycling since
there are odd parity levels above the ground state accessible by electric dipole decay from $|e\rangle$. Nevertheless a closer examination of the odd parity levels below $|e\rangle$ with $15/2\le J\le 19/2$ reveals that almost all possible transitions  have $|\Delta j|>1$ on a single electron or are forbidden due to intercombination spin changes.  The only transition which is allowed for single electron jumps is  $|e\rangle - |4f^{11}5d6s(^1D),J=17/2\rangle$ at a   transition energy of 475 cm$^{-1}$. This transition is very weak due to the $\omega^3$ factor in the expression for the radiative linewidth, with $\omega$ the transition frequency. We can roughly estimate the decay rate using hydrogenic orbitals which gives $\gamma'\sim 4000. ~\rm s^{-1}$. The actual radial matrix elements are unknown so this value is not quantitatively correct but provides some guidance.  To the extent that LS coupling is a good description for the level structure of Ho we expect that the cooling transition with  $|g\rangle$ as an upper level will have small leakage since 
$\gamma'/\gamma\sim 10^{-5}.$

\begin{figure}[!t]
 \centering
   \includegraphics[width=0.5\textwidth]{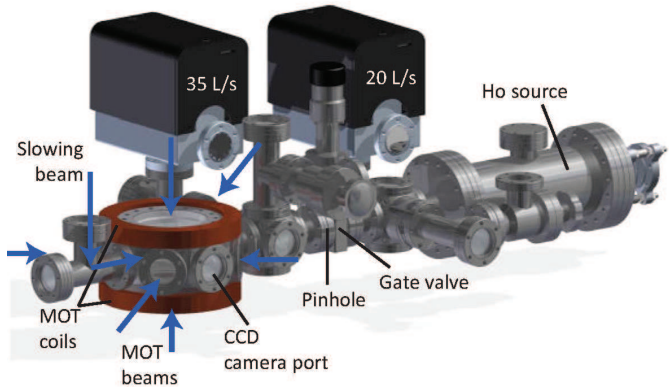}
\vspace{-.6cm}
   \caption{Vacuum and laser cooling setup. The length of the apparatus from end to end is about 1 m. The slowing beam enters the vacuum chamber  through a vertical window and is reflected from a Cu mirror to propagate towards the atomic source.}
   \label{fig.vacuum}
\end{figure}

The experimental apparatus is shown in 
Fig. \ref{fig.vacuum}.  A water cooled effusion cell with Ta crucible operated at $T=1150 ~\rm C$ provides a beam of Ho atoms with a mean velocity of 510 m/s. The atomic beam passes through a 0.25 cm  diameter tube to prevent outflow of any melted Ho from the horizontally oriented effusion cell, and a 0.25  cm diameter aperture for differential pumping before entering the MOT chamber. Two  ion pumps provide a base pressure of $ 10^{-9}~\rm  mbar$ in the  MOT chamber. A pair of electric coils provide a quadrupole magnetic field with a gradient 
of up to 0.4 T/m (vertical axis) and 0.2 T/m (horizontal axis). The cooling beams were arranged in a standard 6 beam $\sigma_+ - \sigma_-$ configuration. The beams had Gaussian waists ($1/e^2$ intensity radius)  of 2.4 mm  and 
total incident power of 40 mW which was doubled by retroreflecting the beams.  The  cooling light was detuned by $\Delta_c\sim -1.5 \gamma$ from the $F_g=11 \rightarrow F_e=12$ cycling transition. Repump light was tuned to $F_g=10 \rightarrow F_e=11$ and  overlapped with all MOT beams. In addition a circularly polarized   slowing beam counterpropagating to the atomic beam was detuned by $\Delta_s=-2\pi\times 320 ~\rm MHz$ from the $F_g=11 \rightarrow F_e=12$ transition. The slowing beam had a power of 140 mW and was focused to a waist of 1. mm. Measurements of the MOT atom number and density were made with an electron multiplying charge coupled device (EMCCD) camera using either fluorescence imaging or absorption imaging with an additional beam tuned to be resonant with the cooling transition. 
All laser beams were derived from a frequency doubled Ti:Sa laser (M$^2$ Solstis with ECD-X) providing up to 1.5 W of 410.5 nm light. The laser was locked to the cooling transition using saturation spectroscopy in a hollow cathode lamp.
 The frequency and power level of the cooling, repump, and slowing light was controlled by acousto-optic modulators. 

The  hyperfine energies shown in Fig. \ref{fig.Holevels}  were calculated from known values  of the $A$ and $B$ 
constants for the ground state\cite{Goodman1962} and measured values for the excited state. We measured the hyperfine constants of the upper level of the cooling transition using modulation transfer spectroscopy in the hollow cathode lamp. Fits to our data 
gave $A=654.9\pm 0.3, ~B=-620\pm 20~\rm MHz$. These values agree well with earlier measurements reported in \cite{Newman2011}.

\begin{figure}[!t]
 \centering
\includegraphics[width=0.5\textwidth]{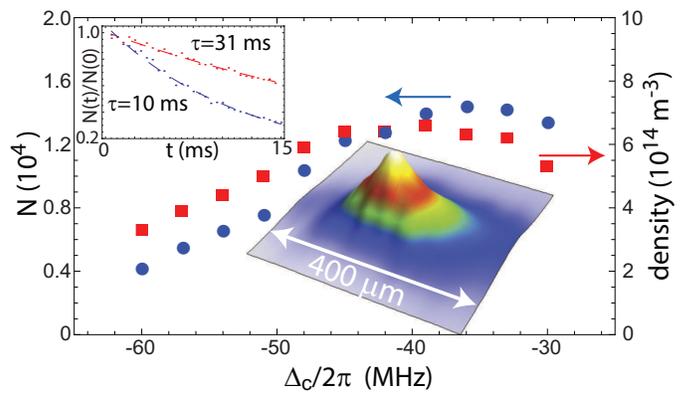}
\vspace{-.8cm}
   \caption{(color online) Number of trapped atoms (circles) and peak density (squares) as a function of cooling light detuning with the slower light turned on and the repump light turned off. The magnetic field gradient was 0.3 T/m.   The insets show an averaged fluorescence image of the trapped atoms from 100 exposures  at -36 MHz detuning and population decay curves with exponential fits at -15 MHz (fast decay) and -45 MHz (slow decay). }
   \label{fig.motnumber}
\end{figure}

With the slowing beam turned on, but no repump light, we achieved a typical MOT population of $N\sim 1.5 \times 10^4$ and atomic density of $n_a\sim 6.5 \times 10^{14} ~\rm m^{-3}$ as shown in Fig. \ref{fig.motnumber}.   Additional data taken with 290 mW of MOT light, beam waists of 1.1 cm, and 38 mW of slower light gave larger MOTs with up to $N=2\times 10^5$ atoms. 

The atom number measurements were calibrated by integrating the detected EMCCD MOT image using the measured camera sensitivity to 410.5 nm light and an integration time of $50~\rm ms$. The measurement relies on knowing the rate of scattered photons per atom 
which we estimated by  the two-level expression\cite{Metcalf1999}
\begin{equation}
r =\gamma\rho_{ee}= \frac{\gamma}{2}\frac{I_T/I_s}{1 + 4\Delta_c^2/\gamma^2 + I_T/I_s }.
\label{eq.rc}
\end{equation}
Here $\rho_{ee}$ is the excited state fraction, $I_T$ is the total intensity of the six MOT beams and  for the saturation intensity we use $I_s= 2.76\times I_{sc}$. Here $I_{sc} = 614.~\rm W/m^2$ is the saturation intensity for the cycling transition $|F_g=11,M=11\rangle \leftrightarrow|F_e=12,M=12\rangle$ and the factor of 
$2.76 = 3 (2\times 11+1)/(2\times 11+3)$ accounts for averaging over   Zeeman substates and the random light polarization in the MOT region. The scattered light was collected with a lens of numerical aperture  0.05 
and imaged onto a calibrated electron multiplying charge coupled device camera which allowed us to deduce the atom number on the basis of the camera photoeletron counts.

When the repumper was turned on tuned to the $F_g=10\rightarrow F_e=11$ transition  the atom number values in Fig. \ref{fig.motnumber} increased by less than 1\%. The negligible influence of the repump light is due to the fact that the cooling light also repumps population in $F_g=10$ much faster than the depumping rate out of $F_g=11$. The depumping rate due to Raman transitions 
via $F_e=10$ or $11$ is  calculated by averaging over $M$ levels and light polarization, and accounting for the branching ratios of the fluorescence decay. We find 
\begin{eqnarray}
r_R &=& \gamma^3 \left[\frac{105}{69938\Delta_{F_e=F_g}^2} 
  + \frac{455}{17904128\Delta_{F_e=F_g-1}^2}\right]\frac{I_T}{I_{sc}}.
\end{eqnarray}
In this expression  $F_g=11$ is the $F$ value for the lower level of the cycling transition, $\Delta_{F_e=F_g}=\Delta_{(F+1)_e,F_e}+\Delta_c,$  $\Delta_{F_e=F_g-1}=\Delta_{(F+1)_e,(F-1)_e}+\Delta_c$ are the detunings of the MOT light from  $F_e=11$ and $F_e=10$, and we have assumed the light is far detuned so that we may replace factors of $(1+4\Delta^2/\gamma^2 + I_T/I_{s})^{-1}$ by $\gamma^2/(4\Delta^2)$. The excited state hyperfine splittings shown in Fig. \ref{fig.Holevels} are $\Delta_{12,11}=2\pi\times 7.67~\rm GHz$ and  $\Delta_{12,10}=2\pi\times 14.79~\rm GHz.$ We find that for $\Delta_c=-2\gamma$, the largest detuning we have used, $r_R = 80 ~\rm s^{-1}$ and $r_R/r = 3.3\times 10^{-6}$. There is also depumping due to leakage to metastable states. The rate for this process  we estimate below to be negligible compared to the Raman rate. The total depumping rate is therefore $< 100 ~\rm s^{-1}$. This rate is balanced by the repumping rate for the Raman process $F_g=10\rightarrow F_e=11\rightarrow F_g=11$. The cooling light is detuned from the $F_g=10\rightarrow F_e=11$ transition by $-610 ~{\rm MHz} + \Delta_c/2\pi\sim 650~\rm MHz$. The resulting repumping rate is about $8\times 10^5~\rm s^{-1}$. 

The ratio of the depumping to repumping rates implies that 
99.9\% of the population is in $F_g=11$. We verified that turning on a separate repumper increased the atom number by less than 
1\%, with the resolution limited by the stability of the MOT  number measurement. Thus, even without a separate repumper, the atoms are pumped with better than 99\% purity into the $F_g=11$ level. Further pumping into a specific Zeeman sublevel as a starting point for quantum state control experiments has not been demonstrated, but should be straightforward using an additional $\pi$ polarized beam tuned to  $F_g=11\rightarrow F_e=11$ for pumping into $M=0$, or a $\sigma_+$ polarized beam tuned to $F_g=11\rightarrow F_e=12$ for pumping 
into $M=11$.

\begin{figure}[!t]
 \centering
   \includegraphics[width=0.5\textwidth]{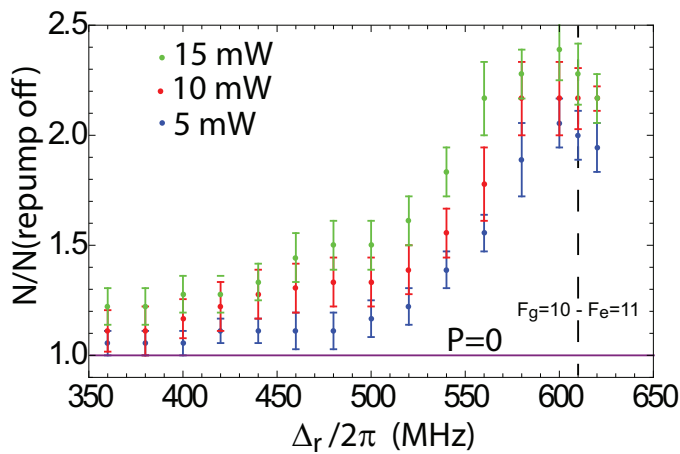}
\vspace{-.8cm}
   \caption{MOT population as a function of  repumper detuning relative to the $F_g=11 \rightarrow F_e=12$ transition,  and repumper power, with the slowing beam off. The population is normalized to the value with no repumper. The data stops at 620 MHz which was the limit of the AOM used for frequency shifting.}
   \label{fig.repump}
\end{figure}

The effect of the repumper is more significant when the MOT is operated without a slowing beam. In this case the MOT captures only a very small fraction of the low velocity tail of the atomic beam and the loading rate is reduced by almost three orders of magnitude, resulting in a very small MOT with $N=40$ atoms, again without a repumper. Turning on the repumper increases $N$ by a factor of up to 2.5, depending on the detuning of the repumper, as shown in Fig. \ref{fig.repump}. The increase in atom number cannot be due to additional repumping since, as explained above, the cooling light acts as its own repumper. We instead interpret the data as being due to the repump light cooling and trapping additional atoms on the $F_g=10\rightarrow F_e=11$ transition. The high temperature atomic source creates a beam which we assume to be uniformly distributed among ground Zeeman states. Since there are 23 Zeeman states in $F_g=11$ and $21$ in $F_g=10$ we expect roughly a factor of 2 increase, which is in reasonable agreement with the data. No such increase is seen when the slower is on since the slower light is only effective for atoms in $F_g=11$ and the additional loading of unslowed $F_g=10$ atoms is negligible.

\begin{figure}[!t]
 \centering
   \includegraphics[width=0.48\textwidth]{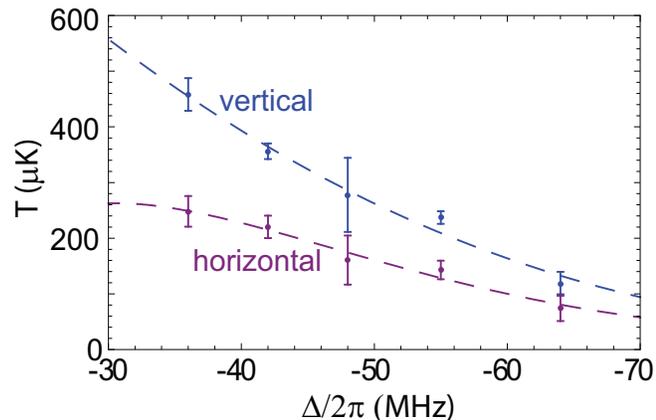}
\vspace{-.8cm}
   \caption{(Color online) MOT temperature with quadrupole magnetic field on versus detuning at saturation parameter $I/I_s=1.1$. 
The temperatures were extracted from fits to  time of flight data for free expansion times up to 15 ms. 
The Doppler temperature is 780 $\mu\rm K$. }
   \label{fig.temp}
\end{figure}

As shown in Fig. \ref{fig.temp} we observe deep sub-Doppler cooling well below $T_D=780~\mu\rm K$  using $\sigma_+ - \sigma_-$ MOT beams. Data were acquired by observing MOT expansion and fitting to Gaussian density profiles.
The theoretical dependence of the molasses temperature with respect to detuning is given by\cite{Dalibard1989}
\begin{equation}
T = T_0 + \frac{\hbar\gamma^2}{2k_B\left|\Delta\right|}\frac{I}{I_s}\left(a+ \frac{b}{1+4\Delta^2/\gamma^2}\right)
\label{eq.TvsD}
\end{equation}
with $a,b$ constants and $T_0$ the low intensity temperature limit. The parameters $a,b$ are known for low angular momentum transitions\cite{Dalibard1989}, but have not been calculated for the $F_g=11\leftrightarrow F_e=12$ Ho cooling transition. Using separate fit parameters for the horizontal and vertical temperatures  Eq.   (\ref{eq.TvsD}) reproduces the observed dependence on detuning quite well. Cooling anisotropy has also been observed in Dy MOTs\cite{Youn2010b} and may be due to the large magnetic moments present in rare earth atoms.    
Due to the near equality  of the ground state $F_g=11$ and excited state $F_e=12$ $g$-factors, 0.82 and 0.83, temperatures reaching ten times below the Doppler limit are achieved with relatively small detunings and with the quadrupole MOT field on during the entire cooling phase.

In order to further study the internal state dynamics we loaded the MOT to steady state with the slowing beam on and then measured the decay lifetime with the slowing beam turned off, which greatly reduces  the loading into the MOT, but the magnetic quadrupole field still on. 
The MOT population decay  can be modeled with the rate equation\cite{Sesko1989} 
\begin{equation}
\frac{dN}{dt}=R_{\rm off}-\Gamma N- \beta N^2.
\label{eq.decay}
\end{equation}
Here
$R_{\rm off}$ is the loading rate with the slowing beam off, and the loss rate is 
 $\Gamma=(1-\rho_{ee})\gamma_{\rm bg}  + \rho_{ee}(\gamma_{\rm la} +\gamma_{\rm ms})$ where $\gamma_{\rm bg}$ accounts for losses due to background collisions with ground state atoms, $\gamma_{\rm la}$ is the  loss rate due to light assisted collisions between optically excited and background  atoms\cite{Bjorkholm1988}, $\gamma_{\rm ms}$ is the loss rate due to scattering into metastable states, and $\beta$ accounts for losses from collisions of trapped atoms.
Leakage into metastable states may lead to atom loss since the metastable states do not interact with the cooling light and the fraction of atoms which are not magnetically trapped will drift away. 
The lowest energy odd parity states with $J=15/2,13/2,11/2,9/2$ are fine structure levels of the ground term. They decay via magnetic dipole transitions which have been calculated in the LS coupling approximation. The resulting lifetimes indicated in Fig. \ref{fig.Holevels} range from 0.02 to 0.24 s. Also the lowest energy even parity states  with $J=17/2,19/2,21/2$ are expected to be long lived, but with lifetimes that are unknown. 

Measurements of the MOT number  decay curve with the slowing beam turned off showed no indication of fast two-body loss ($\beta\approx 0)$. The MOT lifetime was found to range from 10-43 ms as the detuning was scanned from $-5$ to $-50~\rm MHz$. Representative decay curves are shown in Fig. \ref{fig.motnumber}.  Fitting Eq. (\ref{eq.decay}) to the data with 
$\rho_{\rm ee}$ determined from  Eq. (\ref{eq.rc}) we found 
$\gamma_{\rm bg}\simeq 1\pm 1~ {\rm s^{-1}}$ and $\gamma_{\rm ms}+\gamma_{\rm la}\simeq 300.\pm 100. ~{\rm s^{-1}}$.  
The
 value for $\gamma_{\rm la}$ is unknown so the value of $300. ~{\rm s^{-1}}$  should be taken as an upper limit on the 
scattering rate into metastable loss states. 
 This value is much larger than those reported for Yb\cite{Loftus2000} and Tm\cite{Sukachev2010} but is  much smaller than  the value
reported  for   $^{165}$Ho\cite{Hemmerling2013}  ($\gamma_{\rm ms}=1510 ~\rm s^{-1}$). The Ho measurements in \cite{Hemmerling2013} were taken with a much higher background pressure, mainly He, which might contribute to the discrepancy.  
 The  branching ratio to metastable states is in any case very small $\gamma_{\rm ms}/(\gamma+\gamma_{\rm ms}) \sim  10^{-6}$ so that loss out of the cycling transition is dominated by Raman transitions to other ground hyperfine levels with a branching ratio we estimated above to be $\sim 10^{-5}$. 
On the basis of these measurements we conclude that it should be possible to make high fidelity hyperfine state resolved measurements of Ho atoms, which will be  important for qubit experiments.

In conclusion, we demonstrated a magneto-optical trap of Ho atoms. Single frequency molasses light together with a slowing beam is sufficient to load a MOT with as many as $10^5$ atoms at temperatures below $100~\mu\rm K$  from an atomic beam source. The atoms are prepared in the $F_g=11$ level with probability $\sim 99\%$ and fractional leakage rates out of the cooling transition are  $\sim  10^{-5}$.  Future work will explore the use of this cold sample as the starting point for quantum control experiments which take advantage of the large manifold of hyperfine states. 

This work was supported by NSF grant PHY0969883 and the University of Wisconsin graduate school. We are grateful to Jake Covey, Johannes Nipper, and Hannes  Gorniaczyk for contributions at early stages of this experiment and we thank Boerge Hemmerling from the Doyle group for helpful discussions
 concerning measurements of the metastable decay rate.

\bibliographystyle{apsrev4-1}

\end{document}